\documentclass[
pra,aps,
reprint,
a4paper,
superscriptaddress,
longbibliography,
floatfix
]{revtex4-1}
\usepackage{times}
\usepackage{graphicx}
\usepackage{epsfig}
\usepackage{amsfonts}
\usepackage{amsmath}
\usepackage{amssymb}
\usepackage{color}
\usepackage{multirow}
\usepackage[colorlinks=true,linkcolor=blue,citecolor=magenta,urlcolor=blue]{hyperref}

\begin{document}


\title{Quantum correlations with a gap between the sequential and spatial cases}



\author{Zhen-Peng Xu}
\email{zhenpengxu@us.es}
\affiliation{Theoretical Physics Division, Chern Institute of Mathematics,
	Nankai University,
	Tianjin 300071, People's Republic of China}
\affiliation{Departamento de F\'{\i}sica Aplicada II,
	Universidad de Sevilla,
	E-41012 Sevilla, Spain}

\author{Ad\'an~Cabello}
\email{adan@us.es}
\affiliation{%
Departamento de F\'{\i}sica Aplicada II,
Universidad de Sevilla,
E-41012 Sevilla,
Spain}


\begin{abstract}
We address the problem of whether parties who cannot communicate but share nonsignaling quantum correlations between the outcomes of sharp measurements can distinguish, just from the value of a correlation observable, whether their outcomes were produced by sequential compatible measurements on single systems or by measurements on spatially separated subsystems. We show that there are quantum correlations between the outcomes of sequential measurements which cannot be attained with spatially separated systems. We present examples of correlations between spatially separated systems whose quantum maximum tends to the sequential maximum as the number of parties increases and examples of correlations between spatially separated systems whose quantum maximum fails to violate the noncontextual bound while its corresponding sequential version does.
\end{abstract}



\maketitle


\section{Introduction}


Quantum correlations, here defined as the correlations between the outcomes of compatible sharp measurements (i.e., repeatable and only disturbing incompatible measurements \cite{CY14,Kleinmann14}), exhibit many nonclassical features. Quantum correlations between spacelike separated measurements on entangled quantum systems violate local realism as shown by the violation of Bell inequalities \cite{Bell64}. This is known as quantum nonlocality. Quantum correlations between timelike separated compatible sharp measurements on arbitrary quantum states violate noncontextual realism as shown by the quantum state-independent violation of noncontextuality inequalities \cite{Cabello08}. Quantum contextuality is the collective term used to refer to the quantum violations of noncontextuality inequalities (including Bell inequalities) by either single-particle or multiple-particle systems. Noncontextuality inequalities are bounds on linear combinations of probabilities $P(a,\ldots,c|x,\ldots,z)$ of obtaining outcomes $a,\ldots,c$ for compatible measurements $x,\ldots,z$ without making assumptions on how compatibility is achieved.

Quantum correlations between sequential measurements also are referred to as temporal correlations. Quantum temporal correlations are nonclassical in several senses. For example: (i) Their classical simulation requires memory higher than the information-carrying capacity of the quantum system \cite{K11}. (ii) Their classical simulation with systems of a finite number of states requires emission of heat due to Landauer's principle \cite{CGGLW16}. (iii) They outperform their classical counterparts for tasks allowing equal, but limited, communication resources \cite{RC17}. 

\begin{figure}[tb]
	\centering
	\includegraphics[width=8.3cm]{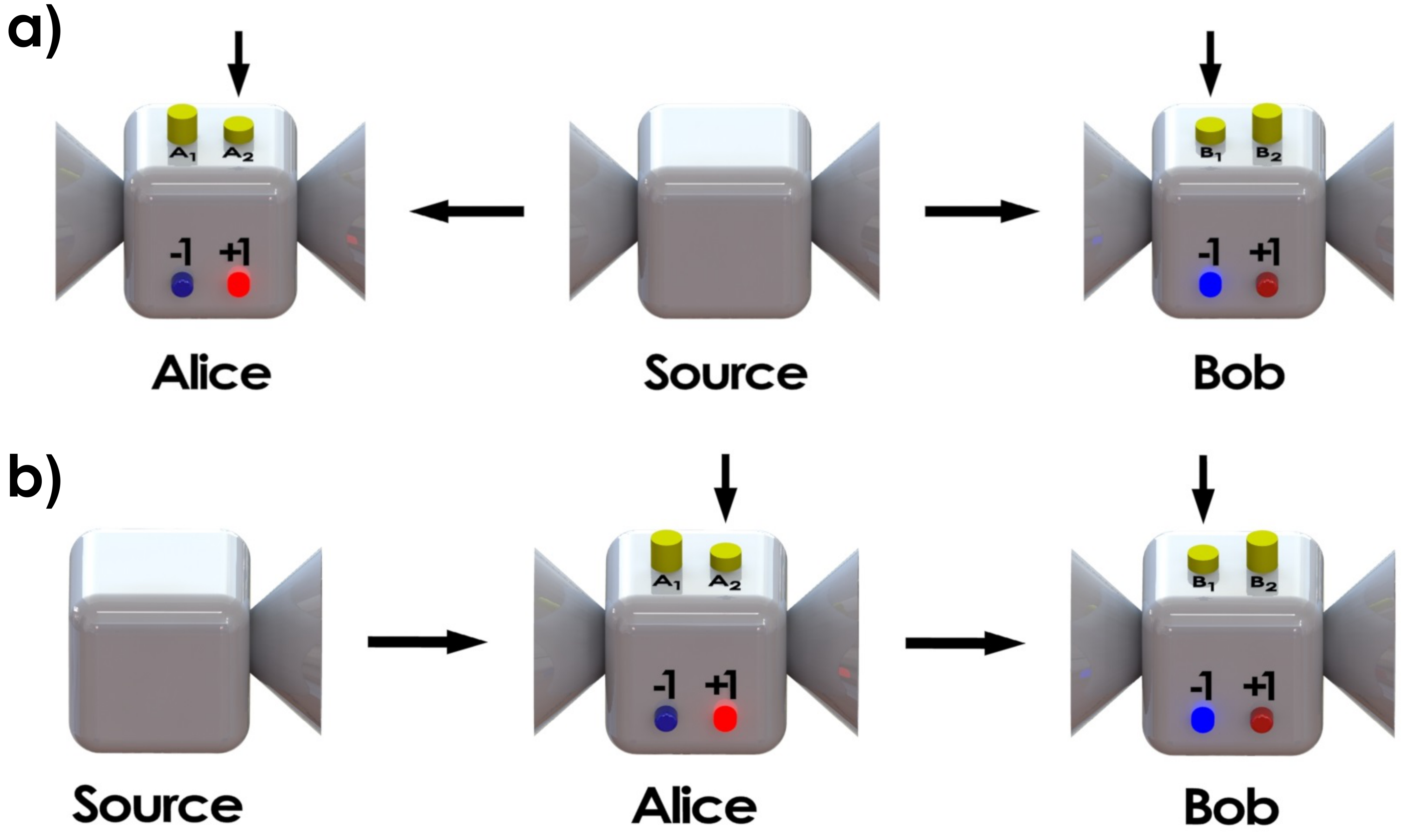}
	\caption{\label{Fig1}(a) Scenario in which two parties, Alice and Bob, perform measurements on two subsystems of a composite system. The measurement setting is indicated by the yellow button that is pressed. The measurement outcome is the light (red or blue) that flashes when the button is pressed. (b) Scenario in which two parties perform sequential compatible measurements on a single system.}
\end{figure}

For illustrating the connection between quantum spacelike separated correlations and quantum sequential correlations between compatible sharp measurements, it has been pointed out \cite{Cabello11,AQB13} that, for many correlation observables including the one in the Clauser-Horne-Shimony-Holt (CHSH) inequality \cite{CHSH69}, the predictions of quantum theory are exactly the same no matter whether: (i) the experiment is performed with spacelike separated measurements, such as in Ref.\ \cite{ADR82}, (ii) with timelike separated measurements on spatially separated subsystems, such as in Ref.\ \cite{FC72}, or (iii) with sequential measurements on single four-level systems, such as in Refs.\ \cite{MWZ00,YLB03}. The typical configuration for experiments of types (i) and (ii) is illustrated in Fig.\ \ref{Fig1}(a), whereas the configuration for experiments of type (iii) is illustrated in Fig.\ \ref{Fig1}(b). In all these cases, the quantum predictions for the correlation operator of CHSH, namely,
\begin{equation}
S_2=\langle A_1 B_1 \rangle + \langle A_1 B_2 \rangle +\langle A_2 B_1 \rangle - \langle A_2 B_2 \rangle
\end{equation}
are the same. Recall that $\langle A_i B_j \rangle$ is the mean value of the product of the results $-1$ or $1$ of $A_i$ and $B_j$, which are measurement settings of Alice and Bob, respectively. In particular, $S_2$ has the same quantum maximum $2 \sqrt{2} \approx 2.828$ in both cases.


However, nonrelativistic quantum theory uses a different mathematical representation for spatially separated measurements as in (i) and (ii) and sequential compatible sharp measurements as in (iii). Measurements on spatially separated systems are represented by operators of the form
\begin{equation}
\label{tensor}
A_i = M_i^A \otimes \openone^B,\;\;\;\;B_j = \openone^A \otimes M_j^B,
\end{equation}
where $\otimes$ denotes the tensor product, $M_m^P$ is an operator in the Hilbert space corresponding to system $P$, and $\openone^P$ is the identity operator in the Hilbert space of system $P$.

On the other hand, sequential compatible sharp measurements on a single system are represented by commuting self-adjoint operators,
\begin{equation}
\label{comm}
[A_i,B_j]=0
\end{equation}
for all $i$ and $j$. Operators satisfying Eqs.\ (\ref{tensor}) automatically satisfy Eq.\ (\ref{comm}). However, Eq.\ (\ref{comm}) can be satisfied in other ways.

As shown in the example of $S_2$, despite these different mathematical representations, in many cases there is no difference between the predictions of quantum theory for spatially separated and sequential correlations. Indeed, e.g., Tsirelson's \cite{Tsirelson80} and Landau's \cite{Landau87} proofs of the quantum maximum of the CHSH inequality use the representation of commuting operators rather than the representation of tensor products. This type of proof, used in some textbooks \cite{Peres02}, is valid if we recall that already CHSH have shown that $2 \sqrt{2}$ can be attained with spacelike separated quantum correlations. In this context, it is interesting a result proven by Tsirelson \cite{SW08} which, for finite dimensional Hilbert spaces, establishes the equivalence of representing local observables on spatially separated systems as in (\ref{tensor}) or as in (\ref{comm}), which is the representation used in algebraic quantum field theory \cite{HK64,Haag96}. However, whether this equivalence also holds for infinite-dimensional Hilbert spaces is still an open problem \cite{SW08,Slofstra16,Slofstra17}.

In any case, in the sequential scenario, and assuming that each party is isolated so that they cannot communicate with each other and despite the fact that there is a system passing from one party to another, if they only have access to the probabilities needed to calculate $S_2$, they cannot ascertain what the other party has performed (i.e., whether the other party already measured or not or which was the measurement the other party performed), as compatibility implies that these probabilities are nonsignaling (i.e., marginal probabilities do not depend on the compatible measurements performed by the other parties). Therefore, the experimental value of $S_2$ does not allow the parties to ascertain whether their outcomes were produced by measurements on spatially separated systems or by sequential sharp measurements on a single system.

In this paper we address the problem of whether there are correlation operators whose value allows the parties who cannot communicate to ascertain whether their outcomes were produced by sequential measurements on a single system or by local measurements on spatially separated systems.


\begin{figure}[tb]
	\centering
	\includegraphics[width=8.6cm]{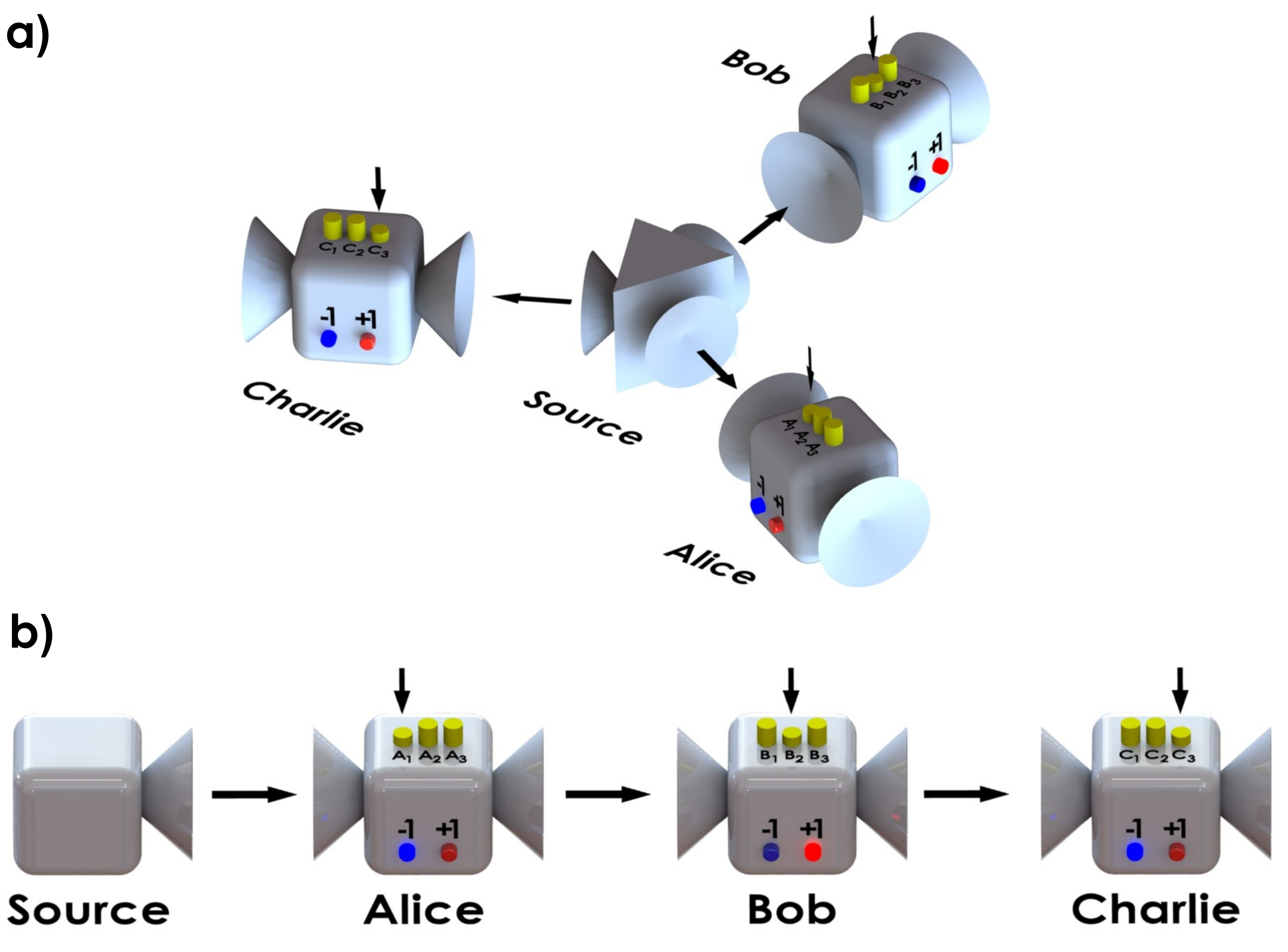}
	\caption{\label{Fig2}(a) Scenario in which three parties perform spatially or spacelike separated measurements on three systems prepared in an entangled state. (b) Scenario in which three parties perform sequential timelike separated compatible measurements on a single quantum system. }
\end{figure}



\section{First example}


Consider the following generalization of the correlation observable of the CHSH inequality for $n \ge 3$ parties, each having $n$ measurement settings $O^1_i, O^2_j,\ldots, O^n_k$ where the super index indicates the party and the subindex indicates her measurement setting with $i,j,\ldots, k=1,\ldots,n$ with possible outcomes $-1$ or $1$,
\begin{equation}\label{Bell}
\begin{split}
S_n \equiv
& \sum_{i=1}^n \langle O^1_i O^2_{i \oplus 1} \cdots O^n_{i \oplus (n-1)} \rangle \\
& + \sum_{i=1}^{n-1} \langle O^1_i O^2_i \cdots O^n_i \rangle - \langle O^1_n O^2_n \cdots O^n_n \rangle,
\end{split}
\end{equation}
where $\oplus$ denotes sum mod $n$.
For local and noncontextual hidden variables the maximum of $S_n$ is $2 (n-1)$. This can be seen as follows. The sets of probabilities consistent with a local or noncontextual hidden variable models are convex polytopes whose vertexes correspond to deterministic assignments for the observables in $S_n$. Since $S_n$ is linear in the mean values, its maximum can always be attained by deterministic assignments to the mean values. Since $S_n$ is a linear combination with weights one of $2n$ mean values and, after any deterministic assignment, the possible values of each of them are $-1$ or $1$, the value of $S_n$ has to be an even number. The only chance for the value to be $2n$ is that the first $2n-1$ terms are $1$ whereas the last term is $-1$. However, this is impossible for local or noncontextual hidden variables, since then
$\prod_{i=1}^n \langle O^1_i O^2_{i \oplus 1} \cdots O^n_{i \oplus (n-1)}\rangle = \prod_{i=1}^n \langle O^1_i O^2_i \cdots O^n_i \rangle$, which means that, when the first $2n-1$ terms are all $1$, then the last term has to be $1$. This proves the maximum for local and noncontextual hidden variable models.

Let us now calculate the quantum maximum of $S_n$. Let us first assume that $O^1_i = M_i^1 \otimes \openone^2 \otimes \cdots \otimes \openone^n$, $O^2_j = \openone^1 \otimes M_j^2 \otimes \cdots \otimes \openone^n,\ldots$, $O^n_k = \openone^1 \otimes \openone^2 \otimes \cdots \otimes M_k^n$. Then, the quantum maximum is
\begin{equation}
\label{eq5}
S_n^{\text{tensor}} = 2n\cos\left(\frac{\pi}{2 n}\right).
\end{equation}
This can be shown as follows. For $n=3$, the method of Navascu\'es {\em et al.} \cite{NPA07} provides an upper bound which, at level $2$ of the hierarchy and up to numerical precision, is equal to $6\cos\left(\frac{\pi}{6}\right) = 3 \sqrt{3} \approx 5.196$. This bound is saturated analytically with projective local measurements on three qubits in an entangled state. Specifically, with the state
$(1+\sqrt{3},1-\sqrt{3},-1+\sqrt{3},1+\sqrt{3})/4 \otimes (1,0)$ and the measurements corresponding to $M^1_1 = F(1/3), M^1_2 = F(2/3), M^1_3 = F(0), M^2_1 = F(1/3), M^2_2 = F(2/3), M^2_3 = F(1), M^3_1=M^3_2=M^3_3=F(1/2)$, where $F(\theta) = \cos(\theta \pi)\sigma_x+\sin(\theta \pi)\sigma_z$. Note that there are only two parties whose alternative measurements contribute to the maximum; the third party always uses the same measurement setting. Something similar happens for $n=4,\ldots,7$. There, the maximum values obtained numerically lead us to conjecture that the quantum maximum is the one given by Eq.\ (\ref{eq5}) and holds for any $n \ge 3$. Then, we notice that this maximum can be attained with only two parties performing alternative measurements. Finally, one can notice that, when we trace out all but two particles, what we have is the bipartite chained Bell inequalities first introduced in Ref.\ \cite{Pearle70} and rediscovered in Ref.\ \cite{BC90}. Since their maximum quantum values are $2n\cos\left(\frac{\pi}{2 n}\right)$ \cite{Wehner06}, this finishes the argument.

Let us now calculate the quantum maximum of $S_n$ when we replace tensor correlations by compatible correlations. Then, if, e.g., we consider the following observables for $n$ parties:
\begin{subequations}
	\begin{align}
	& O^1_1=\sigma_z \otimes \openone,\;\;\;O^2_1=\openone \otimes \sigma_x,
	\ldots,\;\;O^n_1=\sigma_z \otimes \sigma_x, \\
	& O^2_2=\openone \otimes \sigma_z,\;\;\; O^3_2=\sigma_x \otimes \openone,
	\ldots, \;\;O^1_2=\sigma_x \otimes \sigma_z, \\
	& \ldots \nonumber \\
	& O^n_n=\sigma_z \otimes \sigma_z,\;\;O^1_n=\sigma_x \otimes \sigma_x,
	\ldots,\;O^{n-1}_n=\sigma_y \otimes \sigma_y,
	\end{align}
\end{subequations}
where $\sigma_x$, $\sigma_y$, and $\sigma_z$ are the Pauli matrices and all the nondisplayed observables are identities (i.e., they are the trivial observables whose output is always $1$), then $S_n$ achieves its algebraic maximum, i.e.,
\begin{equation}
S_n^{\text{quantum}} = 2n.
\end{equation}
Note that operators in the same row or column commute \cite{Peres90,Mermin90} and, therefore, satisfy the compatibility relations assumed in the definition of $S_n$ given in Eq.\ (\ref{Bell}). 

Therefore, any value of $S_n$ higher than the value for $S_n^{\text{tensor}}$ given by Eq.\ (\ref{eq5}) would allow the parties who cannot communicate to ascertain that they are performing sequential measurements on single systems rather than local measurements on $n$ separated subsystems.

The gap between quantum spatially separated and sequential correlations can be measured by, e.g., $S_n^{\text{quantum}}/S_n^{\text{tensor}}$. As we saw, for $n=2$ (i.e., for the CHSH inequality), there is no gap. The maximum gap occurs for $n=3$, which is the case illustrated in Fig.\ \ref{Fig2}. As the number of parties increases, $S_n^{\text{tensor}}$ tends to $S_n^{\text{quantum}}$. This leads to the question of whether is it possible to find scenarios in which the maximum of the tensor correlations tends to the local maximum as the number of parties increases, whereas there are quantum sequential correlations violating this bound for any number of parties. This is precisely the motivation for the next example.


\begin{figure}[tb]
	\centering
	\includegraphics[width=8cm]{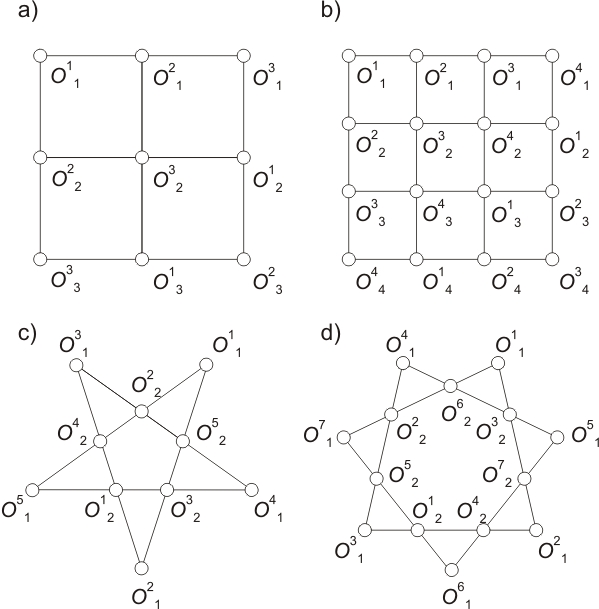}
	\caption{\label{Fig3}(a) Compatibility graph of the observables in $S_3$ defined in Eq.\ (\ref{Bell}). Nodes in the same straight line represent mutually compatible observables. Observables $A_i, B_j, C_k,\ldots$ are Alice's, Bob's, Charlie's,\ldots, respectively. (b) Compatibility graph of the observables in $S_4$. (c) Compatibility graph of the observables in $T_5$ defined in Eq.\ (\ref{Bell2}). (d) Compatibility graph of the observables in $T_7$. Assuming that these observables are, in addition, local observables in a Bell inequality scenario implies the appearance of additional compatibility relations, which, in terms of resources, means the disappearance of incompatibilities and therefore the reduction of the maximum quantum value.}
\end{figure}



\section{Second example}


Consider $2n+1$ parties with $n=2,3,\ldots$ and suppose that party $i$ has two possible measurement settings: $O^i_1$ and $O^i_2$ with $i=1,\ldots,2n+1$. Then, consider the following correlation operator:
\begin{equation}\label{Bell2}
T_{2n+1} \equiv -\sum_{i=1}^{2n+1} \langle O^{i \ominus 1}_2 O^i_1 O^{i\oplus 1}_1 O^{i\oplus 2}_2 \rangle,
\end{equation}
where $\oplus$ and $\ominus$ denote addition and subtraction mod $2n+1$, respectively. Using a similar argument to the one used in the previous section, it can be seen that, for local and noncontextual hidden variable theories, the maximum of $T_{2n+1}$ is $2n-1$.

Using the result in Ref.\ \cite{Masanes03}, which assures that, for the case of $n$ parties with two dichotomic measurement settings each, the quantum maximum for spatially separated measurements on subsystems occurs with projective local measurements on qubits, it can be shown that, for $T_5$, the quantum tensor maximum is 
\begin{equation}
T_5^{\text{tensor}} = 3.340.
\end{equation}
The analytical form of the state needed is too long for displaying it here. However, it can be recovered knowing that the measurement settings are $O^i_1 =\sigma_z$ and $O^i_2=\cos(\pi/4) \sigma_x + \sin(\pi/4) \sigma_z$, for each party $i=1, \ldots, 5$.

However, we have found that, for $n>2$, 
\begin{equation}
T_{2n+1}^{\text{tensor}}=2n-1,
\end{equation}
that is, there is no quantum violation of the hidden-variable bound. 
This is due to the fact that the subspace of operators
that can be represented by tensor products becomes smaller and
smaller as the number of parties increases. Note that this was not the case in the first example where only two parties effectively contributed to the quantum tensor maximum.

Interestingly, for sequential correlations, quantum theory takes the algebraic maximum, namely,
\begin{equation}
T_{2n+1}^{\text{quantum}} = 2n+1.
\end{equation}
This can be proven as follows: First, from the expression of the correlation operator $T_{2n+1}$ in Eq.\ (\ref{Bell2}) we obtain the corresponding compatibility graph, here defined as the graph in which nodes represent observables in $T_{2n+1}$ and observables in the same straight line are compatible. For $n=2,3$, these compatibility graphs are shown in Figs.\ \ref{Fig3}(c) and \ref{Fig3}(d), respectively. 

The Lov\'asz number \cite{CSW14} of a graph $G$ is defined as
\begin{equation}\label{def:Lovasz}
\vartheta(G):= \max \sum_{i\in V(G)} | \langle \psi | v_i \rangle |^2,
\end{equation}
where $V(G)$ is the vertex set of $G$ and the maximum is taken over all sets of unit vectors $\{|v_i \rangle \}$, each of them associated with a node in such a way that nodes in the same straight line are mutually orthogonal vectors and all unit vectors $|\psi\rangle$ in any dimension. 
The Lov\'asz number for the graphs of compatibility associated with $T_{2n+1}$ is
\begin{equation}
\vartheta(T_{2n+1}) = n+\frac{1}{2}.
\end{equation}
Then, we can define the measurement observables in Eq.\ (\ref{Bell2}) as $O^i_j= \openone - 2 |v^i_j\rangle \langle v^i_j|$ and, if we prepare the system in state $|\psi\rangle$, then $T_{2n+1}=2n+1$ since
\begin{equation}
	\begin{split}
		T_{2n+1}^{\text{quantum}} = & -\sum_{i=1}^{2n+1} \left[1 - 2 \langle |v^{i-1}_2\rangle\langle v^{i-1}_2| + |v^{i}_1\rangle\langle v^i_1|\right. \\
		& \left. + |v^{i+1}_1\rangle\langle v^{i+1}_1| + |v^{i+2}_2\rangle\langle v^{i+2}_2| \rangle\right]\\
		= & -(2n+1) + 4 \sum_{i=1}^{2n+1}\sum_{j=1}^2 \langle |v^i_j\rangle\langle v^i_j|\rangle\\
		= & 2n+1.
	\end{split}
\end{equation}
This finishes the proof.


\section{Conclusions}


There are noncontextuality inequalities which can be interpreted both as Bell inequalities involving spatially separated parties acting on composite systems and as noncontextuality inequalities involving parties acting sequentially on single systems. The most famous of these inequalities are the most ancient generalization of the CHSH inequality, that is, the bipartite chained Bell inequality with $n\ge 2$ settings per party \cite{Pearle70,BC90}. In these inequalities, the compatibility graph in the spatially separated case has {\em additional} compatibilities with respect to the compatibility graph of the noncontextuality inequality. However, for these inequalities the quantum maxima are the same in both cases \cite{AQB13}. 

In contrast, here we have shown that there are correlation operators for which the difference between their corresponding spatial and sequential compatibility graphs makes a difference for the predictions of quantum theory. Consequently, the value of these correlation operators can be used to distinguish scenarios such as the one in Fig.\ \ref{Fig3}(a) from scenarios such as the one in Fig.\ \ref{Fig3}(b).

All the examples presented here involve three or more parties. Therefore, an obvious question is whether there are examples with two parties. In principle, we see no reason why not if one has three or more settings per party and three or more outcomes per setting. However, we have not found any example. We leave this problem for future research.

Besides their application to certify that the parties are not performing measurements of spatially like subsystems of a composite system, the examples presented here remind us that quantum correlations between compatible sharp measurements are much richer than those arising in Bell inequality scenarios and that this more general view may be an advantage for understanding quantum correlations from first principles \cite{CSW14,Cabello13,Yan13,ATC14,Cabello15}. An interesting problem for future research is identifying the simplest compatibility graphs for which there is a gap between spatial and sequential correlations, together with the problem of identifying bipartite examples having such a gap. 


\begin{acknowledgments}
	We thank O.\ G{\"u}hne and M.\ Kleinmann for conversations and G.\ Ca\~{n}as for his help with the figures.
	This work was supported by Project No.\ FIS2014-60843-P, ``Advanced Quantum Information'' (MINECO, Spain) with FEDER funds,
	by the FQXi Large Grant ``The Observer Observed: A Bayesian Route to the Reconstruction of Quantum Theory,''
	and by the Project ``Photonic Quantum Information'' (Knut and Alice Wallenberg Foundation, Sweden).
	Z.-P.X.\ was supported by the Natural Science Foundation of China (Grant No.\ 11475089) and the China Scholarship Council.
\end{acknowledgments}





\end{document}